# Periodic transmission peaks in non-periodic disordered one-dimensional photonic structures

Ilka Kriegel[a,b], Francesco Scotognella[a,b,c,*]

[a] *Dipartimento di Fisica, Politecnico di Milano, Piazza Leonardo da Vinci 32, 20133 Milano, Italy*

[b] *Center for Nano Science and Technology@PoliMi, Istituto Italiano di Tecnologia, Via Giovanni Pascoli, 70/3, 20133 Milano, Italy*

[c] *Istituto di Fotonica e Nanotecnologie CNR, Piazza Leonardo da Vinci 32, 20133 Milano, Italy*

\* Corresponding author at: Dipartimento di Fisica, Politecnico di Milano, Piazza Leonardo da Vinci 32, 20133 Milano, Italy. E-mail address: francesco.scotognella@polimi.it (F. Scotognella).

**Abstract**

A better understanding of the optical properties of a device structure characterized by a random arrangement of materials with different dielectric properties at a length scale comparable to the wavelength of light is crucial for the realization of new optical and optoelectronic devices. Here we have studied the light transmission of disordered photonic structures made with two and three different materials, characterized by the same optical thickness. In their transmission spectra a formation of peaks, with a transmission of up to 75%, is evident. The spectral position of such peaks is very regular, which is a result of the constraint that all layers have the same optical thickness. This gives rise to a manifold of applications such as new types of bandpass filters and resonators for distributed feedback lasers.

**Introduction**

The periodic modulation of the dielectric constant, at a length scale comparable to the wavelength of light, is the key property of photonic crystals [1–4]. Such materials are employed for the realization of diverse optical devices, as for example distributed feedback laser [5–7] and optical switches [8,9]. The addition of defects in the periodic alternation, or the realization of completely random sequences, results in disordered photonic structures [10–13]. In the case of one-dimensional disordered photonic structures, very interesting physical phenomena have been theoretically predicted or experimentally observed. In a pioneering work by Faist et al. [14], it was calculated the reflectivity of a GaAs/(GaAs)$_n$(AlAs)$_m$ multilayer in which a layer thickness randomness was introduced, and reflectivity was significant for a wide energy region (wider with respect to the bandgap of the periodic structure). Later, an extension of the band gap with the introduction of disorder in multilayers has been theoretically and experimentally demonstrated [15,16]. Generally, the introduction of disorder results in new optical properties that can be very useful for new types of optical device components.

In this work, we have found the formation of a transmission peak at a precise energy region for disordered photonic structures in which all the layers have the same optical thickness. The photonic structures are comprised of two and three different materials, and for both the formation of transmission peaks with a transmission of about 75% is evident. The periodic repetition of those peaks is a result of the precondition that all layers have the same optical thickness.

**Methods**

We considered periodic and disordered photonic structures made with two (A and B) and three (A, B, and C) different materials. The periodic structure composed of two materials repeated the lattice cell AB, whereas the disordered structure randomly arranged the layers A and B along the multilayer. Similarly, the periodic structure composed of three materials repeated the lattice cell ABC, while the disordered structure randomly arranged the A, B, and C layers along the multilayer. Very important is the condition applied to the thickness of the layers, which requires them to have the same optical thickness of, in this case, $d_i=c/n_i$ nm, with $i$=A, B, C and $n_i$ being the refractive index of the corresponding $i$th layer. In the calculations, we have employed various refractive indexes and different total number of layers and chose the value c to be 310.

For the calculations of the transmission spectra of the periodic and disordered photonic structures, we have employed the transfer matrix method [17,18]. The spectra were calculated with a step of 1 meV.

**Results and Discussion**

In Figure 1a we show the transmission spectra of the periodic photonic structure repeating the unit cell AB for 1000 times (i.e. 1000 unit cells). The refractive index of the layer A and B have been chosen to be $n_A$=1.8 and $n_B$=2. The band gap occurs at 1 eV, and its higher order modes are found at 3 and 5 eV. Due to the condition to keep the same optical thickness among the layers, the peak positions are found at $E_m = \frac{1240}{4 \cdot c} \cdot (2m-1) eV \quad with \; m = 1,2 ...$, with an energy separation of $\Delta E = \frac{1240}{2c} eV$. This finding has been reported in a previous work of ours [19] and demonstrates that the photonic band gap position is independent of the chosen refractive index and dependence solely on c. As predicted by the given formulation, the photonic band gap is found at $E_1$=1 eV and its higher order modes at $E_2$=3 eV and $E_3$=5 eV for the chosen value of c=310, and thus, optical thickness of the layers being ($d_i$=310/$n_i$ nm). As demonstrated already previously [19], the gaps at $2E_1$ and $4E_1$ are not present. In the ordered three layer photonic structure (with the number of layers being t=3) repeating the unit cell ABC with $n_A$=1.8, $n_B$=2, and $n_C$=2.2 for 1500 times, each gap $E_m$ of the two layer photonic structure (band gap and higher order gaps at $E_1$=1 eV, $E_2$=3 eV, and $E_3$=5 eV) is split into two gaps. This is another result from the condition of the layer thickness being $d_i$=c/$n_i$ reported by us in Ref. [19]. The energy position $E_{SG}$ of the split gaps follows the rule of $E_{SG} = E_m \cdot \left(\frac{2 \cdot (1,2...t-1)}{t}\right) eV$, with $E_m$ representing the band gap of the two layer structure and t the number of layers in the unit cell of the ordered photonic structure with more than two layers t=2,3,4,... . See black curves in Figure 1a for two layers and Figure 1b for three layers.

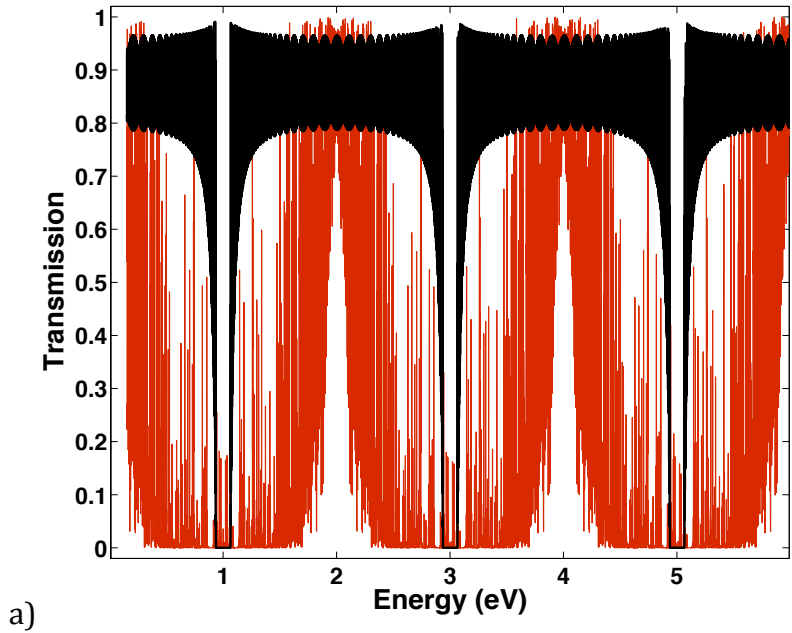

a)

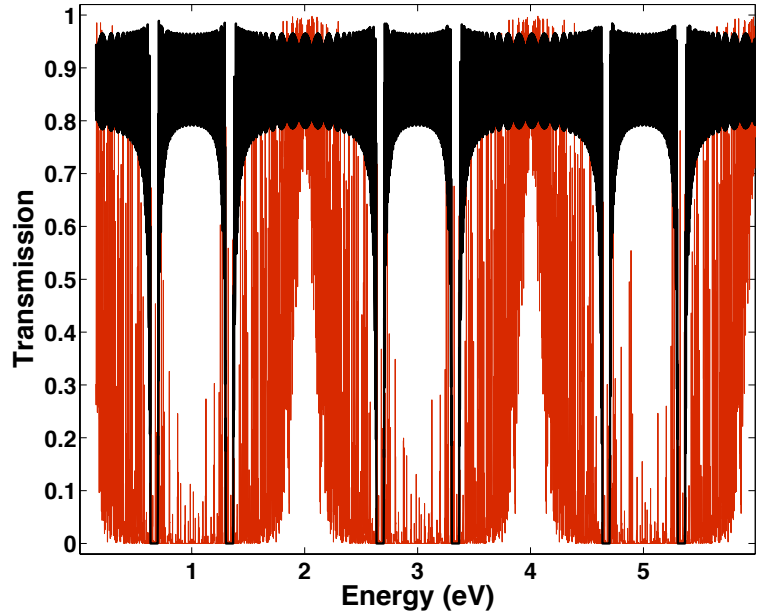

b)

**Figure 1.** a) Periodic three-material photonic crystals (black curves) and random two-material photonic structures (orange curves). Both are made with 1000 layers. b) Periodic three-material photonic crystals (black curves) and random three-material photonic structures (orange curves). Both are made with 1500 layers.

The calculation of the transmission properties of photonic crystal structures with a disordered arrangement of the layers, is based on a random distribution of the layers AB or ABC for a number of 1000 and 1500 layers, respectively. A similar layer thickness and refractive index as given above is used. The transmission spectra of the disordered structure for two and three materials are the orange curves shown in Figure 1a and b, respectively. Interestingly, the completely disordered structures show very regular transmission peaks at 2 and 4 eV, which corresponds to the missing bandgaps at $2E_1$ and $4E_1$, i.e. following $E_n = \frac{1240}{4 \cdot c} \cdot 2n \, eV \, with \, n = 1,2,3 \ldots$. Such transparency windows show a high transmission above 75%. Notably, the transmission spectra of the disordered structure seem to be unaffected by the number of layers with different refractive index taken into account.

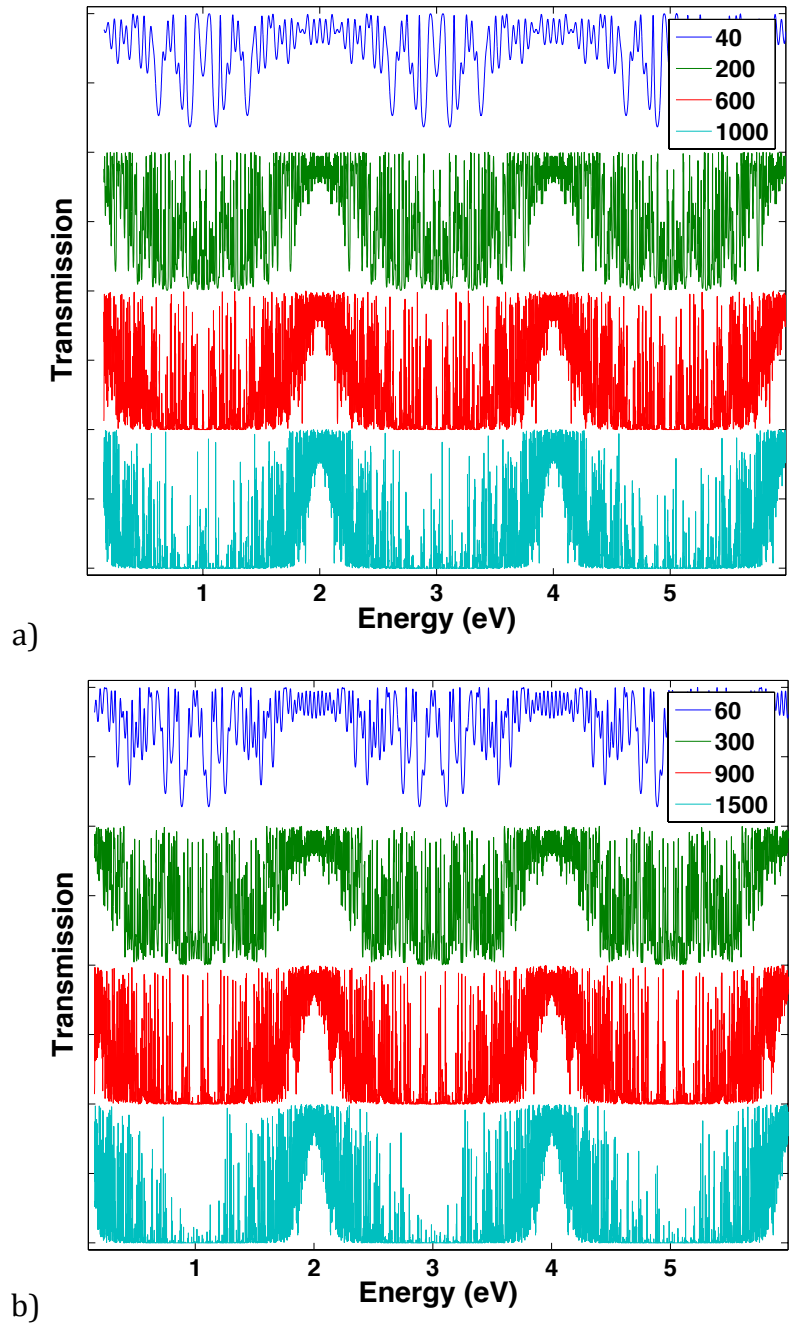

**Figure 2.** a) Random two-material photonic structures made by 40, 200, 600 and 1000 layers; b) Random three-material photonic structures made by 60, 300, 900 and 1500 layers.

To explain the origin of this very characteristic spectra of the disordered structures, we show their behaviour as a function of the number of layers. In Figure 2 we show the spectra for 20, 100, 300, and 500 unit cells, i.e. 40, 200, 600, and 1000 layers for the two-material structures (Figure 2a) and 60, 300, 900 and 1500 layers for the three-material ones (Figure 2b). We observe that with increasing number of layers the transmission minima centred at 1, 3, and 5 eV become more intense and more defined, creating a wide opaque window, in both cases.

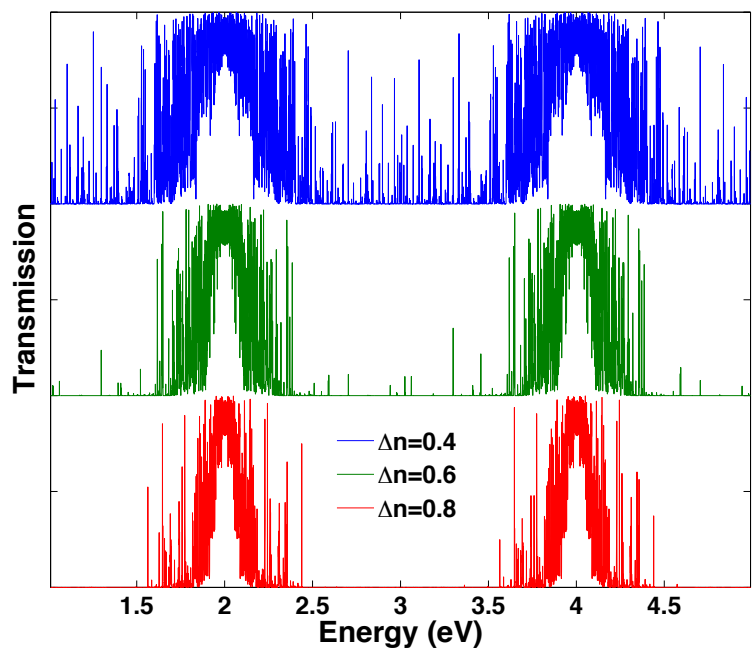

a)

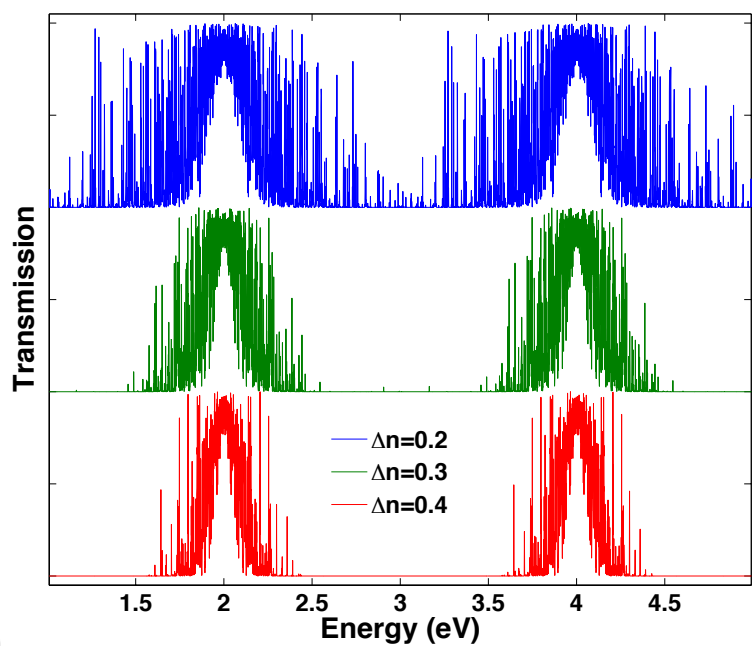

b)

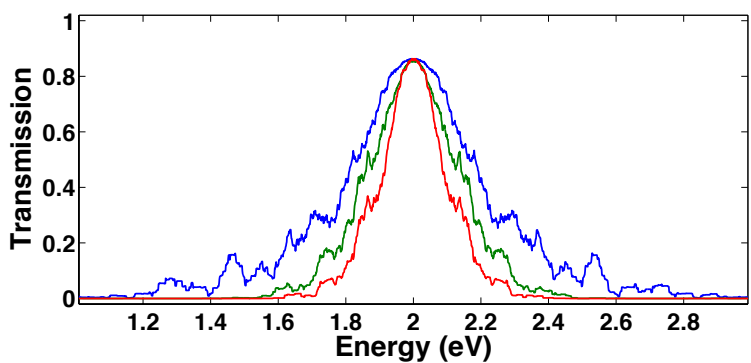

c)

**Figure 3.** a) Transmission spectra for two-material disordered photonic structures with 1000 layers for different refractive index ratio; b) Transmission spectra for three-material disordered photonic structures with 1500 layers for different refractive index ratio. c) Smooth of the transmission spectra in b) in the spectral region 1-3 eV (span of the moving average 50).

We further investigated the dependence of the transmission spectra of our disordered structures on the refractive index contrast. Therefore we calculated the transmission spectra for a varying refractive index contrast $\Delta n$ given in Figure 3 for the two-material and the three-material multilayers (Figure 3a and b, respectively). The centre refractive index in all cases is n=2. Thus, for a $\Delta n$=0.4 in the two layer structure $n_A$=1.8, $n_B$=2.2, while for the three-material one, a $\Delta n$=0.2 means that $n_A$=1.8, $n_B$=2, and $n_C$=2.2. The increase of the refractive index ratio, leads to a more intense light reflection in the opaque spectral regions, with less transmission spikes. Furthermore, as shown in Figure 3c, the width of the transmission peaks decreases (half width at half maximum, HWHM, of 200, 150, and 90 meV for $\Delta n$=0.2, 0.3, and 0.4, respectively). We remark here that the smoothed transmission peaks between the two material system with $\Delta n$=0.4, 0.6, and 0.8 and three material system with $\Delta n$=0.2, 0.3, and 0.4 are very similar, indicating that the main role is here played by the maximum refractive index difference, i.e. by the difference between the highest and lowest refractive index in the structure.

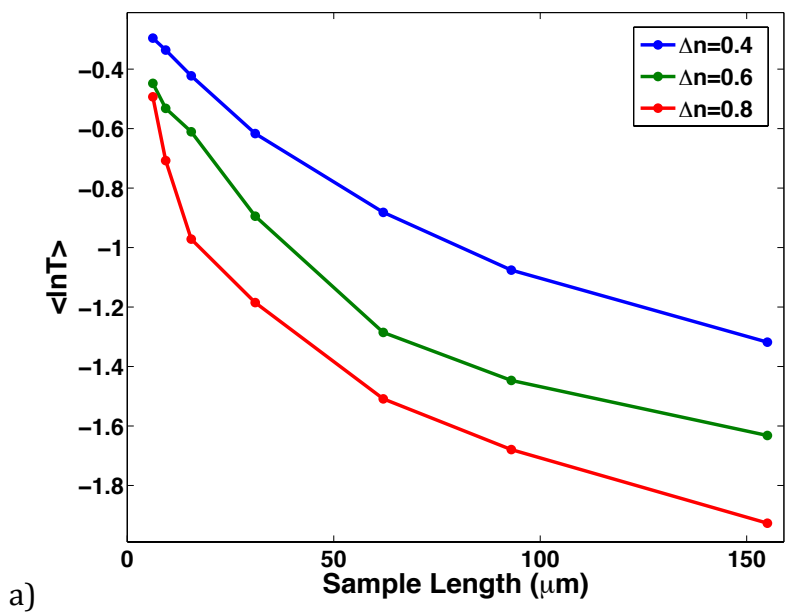

a)

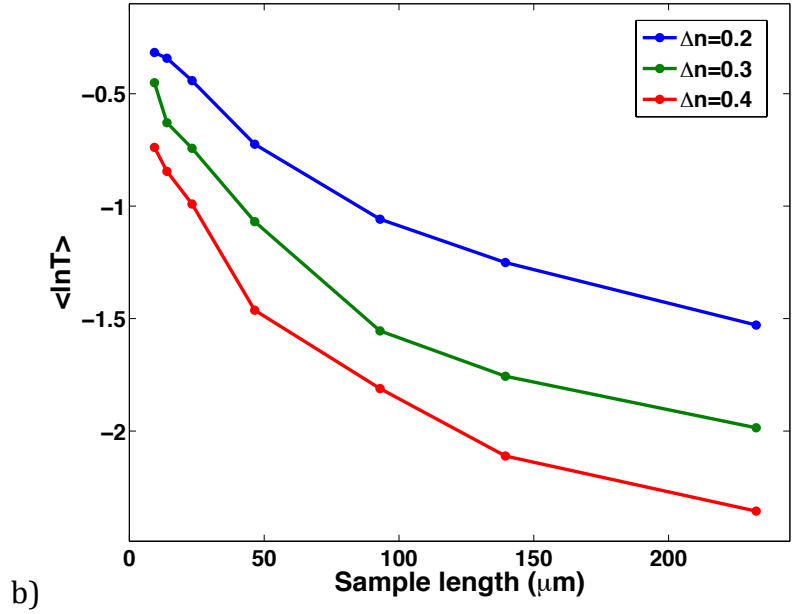

b)

**Figure 4.** Logarithm of the transmission, of a (a) two-material and (b) three-material disordered photonic structure, as a function of the sample length for different refractive index ratio.

We were interested to study the light localization in three-material disordered photonic structures and in Figure 4b we showed the logarithm of the transmission as a function of the sample length for different refractive index ratio. We observed that the slope of *<lnT>*, compared to the one of the two-material structure (Figure 4a) is less intense. This means that in three-material random multilayers a mild localization of light occurs, compared to light localization in the two-material structures, in which an Anderson regime has been observed [20].

**Conclusions**

In this study we have studied the light transmission characteristics of disordered photonic structures made with two and three different materials, in which all the layers have the same optical thickness. In their transmission spectra a formation of peaks, with a transmission of about 75%, is evident. The spectral position of such peaks is very regular and follows $E_n = \frac{1240}{4 \cdot c} \cdot 2n \; eV \; with \; n = 1,2,3 \ldots$. Choosing a value *c*=310 and thus a layer thickness of 310/$n_i$ nm, the peaks occur at integer multiples of 2 eV. The realization of these kind of multilayers are very interesting for the optical bandpass filtering or for laser resonators.

**Acknowledgement**

This work was also supported by Fondazione Cariplo through the project EDONHIST (Grant no. 2012-0844) and Italian Ministry of University and Research (project PRIN 2010-2011 "DSSCX", Contract 20104XET32).